\documentstyle[prd,preprint,aps,epsfig,floats]{revtex}

\begin{document}

\tightenlines
\input epsf.tex
\def\DESepsf(#1 width #2){\epsfxsize=#2 \epsfbox{#1}}
\draft
\thispagestyle{empty}
\preprint{\vbox{ \hbox{UMD-PP-03-004}}}

\title{{\Large\bf Testing Leptonic $SU(2)$ Horizontal Symmetry Using
Neutrino Mixings }}
\author{{\bf R. Kuchimanchi} and {\bf R.N. Mohapatra}}

\address{Department of Physics, University of Maryland, College Park, MD-
20742}
\date{July, 2002}

\maketitle

\begin{abstract}
After some preliminary arguments suggesting that neutrino
mixings with inverted mass pattern may be easier to understand
within the framework of a local horizontal
symmetry $SU(2)_H$ acting on leptons, we construct a
specific extension of the minimal supersymmetric standard model that
implements the idea and analyze its predictions. We show that the
horizontal
symmetry leads to an experimentally testable relation between the neutrino
parameters $U_{e3}$
and the ratio of solar and atmospheric mass difference squared i.e.
$U^2_{e3} cos 2\theta_{\odot} = \frac{\Delta m^2_{\odot}}{ 2 \Delta m^2_A}
+ O(U^4_{e3}, (m_e/m_\mu)^2)$. Taking the solar neutrino parameters
inferred
from present data at 99.7\% confidence level, the above relation leads to
a lower bound on $U_{e3}\geq 0.08$ and an allowed region in the $U_{e3}$
and  $\frac{\Delta m^2_{\odot}}{ 2 \Delta m^2_A}$ space which can be
tested in proposed long baseline experiments.
\end{abstract}

\section{Introduction}
As the outline of the neutrino mixings pattern is beginning to emerge
from recent solar and atmospheric neutrino experiments, understanding the 
neutrino mass matrix has become one of the central problems in theoretical
particle
physics. On the phenomenological side, while the mixings responsible for
both solar and atmospheric neutrino oscillations seem to be fairly
large (unlike quark mixings)\cite{bahcall}, the pattern of masses seem to
remain
undetermined. Three generic patterns that can be considered as equally
acceptable at present are masses (i) with normal hierarchy i.e $m_1 \ll
m_2 \ll m_3$; (ii) inverted hierarchy i.e. $m_1 \simeq -m_2 \gg m_3$ and
(iii) degenerate i.e. $m_1 \simeq m_2 \simeq m_3 $. Once some of the
contemplated long baseline neutrino experiments\cite{lbl} and high
precision searches for neutrinoless double beta decay\cite{dbeta} are
carried out\footnote{If the recent reports of a
positive signal for $\beta\beta_{0\nu}$\cite{klap} are confirmed, the
degenerate mass pattern\cite{caldwell} will be picked as the unique choice
and models of the type discussed here will be disfavored.}, the true
mass pattern will be revealed. From a theoretical
point of view, each pattern could be an indication of a different symmetry
of physics beyond the standard model. Therefore, before those
experiments are carried out, it is of interest in our
opinion to explore the symmetry approach to understanding neutrino masses
and isolate their tests. Combination of the future experimental results
and the theoretical explorations can then decide the nature of physics
beyond the standard model.

The key issues that need to be understood are: (i) the large atmospheric
and solar mixing angles; (ii) the smallness of the ratio $\Delta
m^2_{\odot}/\Delta m^2_A$ and (iii) the smallness of the mixing $U_{e3}$.

One symmetry that predicts in zeroth order the correct mixing pattern
i.e. large solar and atmospheric angles and zero $U_{e3}$ is the
combination of the three leptonic symmetries of the standard model
i.e. $L_e-L_{\mu}-L_{\tau}$\cite{emutau}; it picks the inverted
hierarchy pattern and an exact bimaximal mixing\cite{bimax}. However in
the symmetry limit it predicts that $\Delta m^2_{\odot}=0$ while
$\Delta m^2_A$ is predicted to be nonzero. This, therefore raises the
possibility that, one
may be able to understand the second puzzle in such models.
In fact if one includes small breakings of the $L_e-L_\mu-L_\tau$
symmetry either radiatively\cite{babu} or otherwise, it leads to quite
interesting
and testable neutrino mixing patterns. Because of this a great deal of
attention has
recently been focussed on it\cite{emutau}.

In order to have a deeper theoretical understanding of the inverted
neutrino mass pattern with near bimaximal mixing or the
$L_e-L_\mu-L_\tau$ symmetry, one approach would be to 
study it within a seesaw framework
where the smallness of the neutrino masses is understood in a very simple
manner\cite{seesaw}. It appears that the most convenient way to arrive at
the inverted pattern
with two large mixings in a seesaw framework is to work with two heavy
right handed neutrinos rather than three as is dictated by quark lepton
symmetry\cite{lavoura}. In a recent paper\cite{kuchi}, we pointed out that
if the standard model is extended by the inclusion of an
$SU(2)_H$\cite{hor}
symmetry acting on two lepton families, freedom from global anomalies 
require that there be two right handed neutrinos at the scale where
$SU(2)_H$ symmetry is broken.
We further showed\cite{kuchi} that (i) the presence of the $SU(2)_H$
symmetry
also helps in the understanding of the near bimaximal mixing pattern; (ii)
the smallness of $\Delta m^2_{\odot}/\Delta m^2_A$\cite{ma} is associated 
with a symmetry related to the horizontal symmetry.
Needless to say that while it may appear that theory does not have quark
lepton symmetry, it could be easily
restored by including the  third right handed neutrino and making it
heavier than the seesaw scale. In this case the low energy theory near the
horizontal symmetry breaking scale looks effectively like a theory with
two right handed neutrinos.

It is the purpose of this paper to analyze this class of models in more
detail and point out that in a specific supersymmetric
realization of the model there is an experimentally interesting
 relation between the $U_{e3}$ parameter, the ratio $\Delta
m^2_{\odot}/\Delta m^2_A$ and the solar mixing angle
$\theta_{\odot}$ i.e.
$sin\theta_{\odot}$ as follows: $U^2_{e3} cos 2\theta_{\odot} =
\frac{\Delta m^2_{\odot}}{2 \Delta m^2_A} + O(U^4_{e3},
(m_e/m_\mu)^2)$. Apart
from being
experimentally testable, this relation also provides a natural explanation
of why the $\Delta m^2_{\odot}$ is so much smaller than $\Delta m^2_A$.

We have organized this paper as follows: in section 2, we give arguments
to suggest that within a seesaw framework for neutrino masses, normal or
inverted hierarchical patterns prefer that one of the right
handed neutrinos is much heavier than the others. In section 3, we
present the basic ingredients of the model which is an $SU(2)_H$
extension of the minimal supersymmetric standard model (MSSM); in
sec. 4, we discuss the predictions for
neutrino mixings in  the model and derive the main result of the paper,
which is the  relation between the
neutrino oscillation
parameters discussed above. Section 5 is devoted to some additional
comments and in sec. 6, we give a summary of
our results and conclusions.

\section{Why $SU(2)_H$ ?}
In this section, we will argue that for the normal or inverted hierarchy
case, it is quite possible that two of the right handed neutrinos are
lighter than the third one. This can be seen as follows. Let us assume
that the smallness of the neutrino masses owes its origin to the seesaw
mechanism\cite{seesaw}:
\begin{eqnarray}
{\cal M}_{\nu}~=~-M_D M^{-1}_R M^T_D
\end{eqnarray}
Inverting this relation with the assumption that the Dirac mass matrix
is diagonal, one can express $M_R$ in terms of the neutrino mixing
matrix elements and the neutrino masses $m_i$ and one has:
\begin{eqnarray}
{\cal M}_{R, \alpha \beta}~=~ m_{D,\alpha}\mu^{-1}_{\alpha
\beta}m_{D,\beta}
\end{eqnarray}
with
\begin{eqnarray}
\mu^{-1}_{\alpha \beta}~=~\sum_i U_{\alpha i} U_{\beta i}m^{-1}_i
\end{eqnarray}
where ${\bf U}$ is the neutrino mixing matrix.
Now observe that for both the normal and inverted hierarchy case, the
lightest neutrino could have mass even equal to zero. Clearly as its mass
gets closer to zero, the RH neutrino matrix takes the factorized form
\begin{eqnarray}
M_R~=~m^{-1}_1 | 1 \rangle \langle 1|
\end{eqnarray}
where $| 1 \rangle = (U_{11}m^2_{D1}, U_{21}m_{D1}m_{D2},
U_{31}m_{D1}m_{D3})$
and clearly the smaller $m_1$ is, the
heavier the heaviest right handed neutrino becomes.
On the other hand the masses of the other two RH neutrinos are not free
since they are linked to observed $\Delta m^2_{\odot}$ and $\Delta m^2_A$.
In this sense, we see that for both the normal and the inverted hierarchy
case, it is quite likely that there is a separation of the RH neutrino
levels. In fact, in the case of inverted hierarchy, the two ``heavy''
left handed neutrinos are nearly degenerate, the two lighter RH neutrinos
are likely to be very close in mass, which then makes the case for a
symmetry associated with it. In ref.\cite{kuchi} we argued that the
relevant symmetry is $SU(2)_H$ symmetry, which by group theory argument
alone puts two of the RH neutrinos lighter than the third one.
As discussed in ref.\cite{kuchi}, this happens because, when the local
horizontal symmetry acts on
the charged right handed leptons, freedom from global anomalies
indeed requires that there be two RH neutrinos transforming as a doublet
under $SU(2)_H$. Their mass after $SU(2)_H$ symmetry breaking then would
be of order of the horizontal symmetry breaking scale. The third RH
neutrino being unconstrained by this symmetry would have a much higher
mass.

\bigskip

\section{Details of the model and mass matrices for leptons}
Our model is based on the gauge group $G_{STD}\times SU(2)_H$ with 
supersymmetry.
In Table I the assignment of the leptons and Higgs superfields
under the gauge group $ SU(3)_c\times SU(2)_L\times U(1)_Y \times SU(2)_H
\equiv G_{STD}\times SU(2)_H$ is given.

\begin{center}
{\bf Table I}
\end{center}

\begin{center}

\begin{tabular}{|c||c|}\hline
Particles &  $G_{STD}\times SU(2)_H$\\
    & Quantum numbers \\ \hline\hline
$\Psi \equiv (L_e, L_{\mu})$ & (1,2,-1,2) \\ \hline
$L_{\tau}$  &  (1,2,-1,1) \\ \hline
$E^c \equiv (\mu^c, -e^c)$ & (1,1,-2, 2) \\ \hline
$\tau^c$ & (1,1,-2, 1)\\ \hline
$N^c\equiv (\nu^c_{\mu}, -\nu^c_{e})$ & (1,1,0,2) \\ \hline
$\nu^c_{\tau}$ & (1,1,0,1) \\ \hline
$\chi_H\equiv \left(\begin{array}{cc} \chi_{1} & \chi_2 \\
\end{array}\right) $ & (1, 1, 0, 2) \\ \hline
$\bar{\chi}_H~\equiv (\bar{\chi}_1, \bar{\chi}_2)$ & (1,1,0,2) \\ \hline
$H_u$ & (1,2,1,1)\\ \hline
$H_d$ & (1,2,-1,1) \\ \hline
$\Delta_H$ & (1,1,0,3)\\ \hline\hline
\end{tabular}

\end{center}
\bigskip

\noindent{\bf Table caption}: Representation content of the
various fields in the model under the gauge group $G_{STD}\times SU(2)_H$.

\bigskip
Here $L_{e,\mu,\tau}$ denote the left handed lepton doublet
superfields.  The quarks can transform as singlets or doublets of
$SU(2)_H$ and are not mentioned since it does not mix affect the
lepton masses which is the main focus of this paper.
We arrange the Higgs potential in such a way that the $SU(2)_H$ symmetry
is broken by $<\chi_1>= v_{H1}; <\chi_2>=v_{H2}$ and $<\Delta_{H,3}>=
v'_H$, where $v_H,v'_H \gg v_{wk}$. Note that we have used the $SU(2)_H$
symmetry to align the $\Delta_H$ vev along the $I_{H,3}$ direction. At the
weak scale,
the neutral components of the fields $H_u$ and $H_d$ acquire
nonzero vev's and break the standard model symmetry down to $SU(3)_c\times
U(1)_{em}$. We denote these vev's as follows: $<H^0_{u}>
=\kappa_{0}$ and $<H^0_d>=\kappa_0 cot \beta$ ; Clearly $\kappa_0$ is
expectd to have values in few to 100 GeV range. All the vev's and couplings are
taken to be real.

Note that $<\Delta_H>\neq 0$ breaks the $SU(2)_H$ group down to the
$U(1)_{L_e-L_{\mu}}$ group which is 
further broken down by the $\chi_H$ vev. Since the renormalizable
Yukawa interactions do not involve the $\chi_H$ field, this symmetry
($L_e-L_{\mu}$) is also reflected in the right handed neutrino mass
matrix and plays a role in leading to the bimaximal mixing pattern.

To study the pattern of neutrino masses and mixings, let us first
note that if we included $\nu_{\tau R}$ in the theory, a bare mass for the
$\nu_{\tau, R}$ field is allowed at the
tree level unconstrained by any symmetries. This mass can therefore
be arbitrarily large and $\nu_{\tau, R}$ will decouple from
the low energy spectrum. We will work in this limit of decoupled
$\nu_{\tau R}$ and write
down the gauge invariant Yukawa superpotential involving the remaining
leptonic fields.
\begin{eqnarray}
{\cal W}_Y~=~h_1 (L_eH_u\nu^c_e+L_\mu H_u\nu^c_\mu)
+h_0 L_\tau(\nu^c_\mu \chi_2 + \nu^c_e\chi_1)H_u/M\\ \nonumber
-if N^{c T}\tau_2{\bf \tau \cdot \Delta_H}N^c 
+\frac{h'_1}{M}(L_e\chi_2-L_\mu \chi_1)H_d\tau^c \\
\nonumber
+\frac{h'_4}{M} L_\tau H_d(\mu^c\chi_2+e^c\chi_1) +
{h'_3}L_{\tau}H_d \tau^c
+h'_2 (L_ee^c+L_\mu \mu^c)H_d 
\end{eqnarray}
$<\Delta^0_H>= v'_H$ directly leads to the $L_e-L_{\mu}$ invariant
  $\nu_{eR}-\nu_{\mu R}$ mass matrix at the seesaw scale.
The $\chi_H$ vev contributes to this mass matrix only through
nonrenormalizable operators and we assume those contributions to be 
negligible. We also do not include any term where $\Delta_H$ and
$\bar{\chi}_H$ couple to light fields. Since in supersymmetric theories,
the superpotential does not receive any loop induced corrections due to
the nonrenormalization theorem, conclusions derived on the basis of the
above potential are stable under radiative corrections. 
It must however be noted that even if such couplings were allowed, there
would be no change in the predictions since the effects would be small.
Similarly there will also be some small contributions from
the $\nu_{\tau R}$ sector if we did not decouple it completely. We ignore
these contributions in our analysis. Further, we define
$\kappa_{1,2}= \frac{<\chi_{1,2}>\kappa_0}{M}$ taken to of order 10 GeV
or so.

To study neutrino mixings, we write
down the $5\times 5$ seesaw matrix for neutrinos:
\begin{eqnarray}
M_{\nu_L,\nu_R}~=~\left(\begin{array}{ccccc} 0 & 0 & 0 & h_0\kappa_0 & 0\\
0 & 0 & 0 & 0 & h_0\kappa_0\\ 0 & 0 & 0 & h_1\kappa_1 & h_1 \kappa_2 \\
h_0\kappa_0 & 0 & h_1\kappa_1 & 0 & fv'_H \\ 0 & h_0\kappa_0 & h_1\kappa_2
& fv'_H & 0 \end{array} \right)
\end{eqnarray}
After seesaw diagonalization, it leads to the light neutrino mass matrix
of the form:
\begin{eqnarray}
{\cal M}_{\nu}~=~-M_D M^{-1}_R M^T_D
\end{eqnarray}
where $M_D~=~\left(\begin{array}{cc} h_0\kappa_0 & 0 \\ 0 & h_0\kappa_0\\
h_1\kappa_1 & h_1\kappa_2
\end{array}\right)$; $M^{-1}_R~=~\frac{1}{fv'_H}\left(\begin{array}{cc} 0
&
1\\1 & 0 \end{array}\right)$. The resulting light Majorana neutrino mass
matrix ${\cal M}_{\nu}$ is given by:
\begin{eqnarray}
{\cal M}_{\nu}~=~-\frac{1}{fv'_H}\left(\begin{array}{ccc} 0 &
(h_0\kappa_0)^2 & h_0h_1\kappa_0\kappa_2\\ (h_0\kappa_0)^2 & 0 &
h_0h_1\kappa_0\kappa_1 \\ h_0h_1\kappa_0\kappa_2 & h_0h_1 \kappa_0\kappa_1
& 2h^2_1\kappa_1\kappa_2 \end{array}\right)
\end{eqnarray}
To get the physical neutrino mixings, we also need the charged lepton mass
matrix defined by $\bar{\psi}_L {\cal M}_\ell \psi_R$. This is given in
our model by:
\begin{eqnarray}
{\cal M}_{\ell}~=~cot\beta \left(\begin{array}{ccc} h'_2\kappa_0
& 0
&-h'_1\kappa_2
\\ 0 & h'_2\kappa_0 & h'_1\kappa_1 \\ h'_4\kappa_1
& h'_4\kappa_2 &
h'_3\kappa_0 \end{array}\right)
\end{eqnarray}
In order to study physical neutrino mixings, we must diagonalize the
${\cal M}_{\nu}$ and $M_\ell$ matrices. We discuss this in the next
section.

\section{A relation between Neutrino mixings and smallness of $\Delta
m^2_{\odot}/\Delta m^2_A$}

In order to discuss the physical neutrino mixings, we need to work in a
basis where the charged lepton mass matrix is diagonal.
Defining the matrices that diagonalize the charged lepton mass matrix as 
$D_\ell = U^{(L)}_{\ell} M_{\ell} U^{(R)\dagger}_{\ell}$, we get 
\begin{eqnarray}
U^{(L)}_{\ell}=\left(\begin{array}{ccc}
s_1 & c_1 & 0\\c_\beta c_1 & -c_{\beta}s_1 & s_\beta \\
-s_\beta c_1 & s_\beta s_1 & c_\beta \end{array}\right)
\end{eqnarray}
where $tan 2\beta \simeq 2\sqrt{\frac{h'_1}{h'_4}} \sqrt
{m_{\mu}/m_{\tau}}$ and $sin \theta_1 \equiv s_1 =
\frac{\kappa_1}{\sqrt{\kappa_1^2+\kappa_2^2}}$. This matrix receives
small corrections of order
$m_e/m_\mu$, which are not important for our considerations.

In order to discuss neutrino mixings, we write down the
orthogonal matrix $U_{\nu}$ that  diagonalizes the ${\cal M}_{\nu}$
for $\kappa_1\neq 0$. Defining two angles $\theta_{1,2}$:
\begin{eqnarray}
sin \theta_1 \equiv s_1 = \frac{\kappa_1}{\sqrt{\kappa_1^2+\kappa_2^2}}\\
\nonumber
sin \theta_2 \equiv s_2 = \frac{h_0\kappa_0}{\sqrt{h^2_0\kappa^2_0+
h^2_1(\kappa^2_1 +\kappa^2_2)}} \\ \nonumber
\end{eqnarray}
the neutrino mass matrix can be written in terms of these angles as:
\begin{eqnarray}
{\cal M}_{\nu}~=~\sqrt{\Delta m^2_A} \left(\begin{array}{ccc} 0 & s^2_2 &
c_2s_2c_1\\
s^2_2 & 0 & c_2s_2s_1 \\ c_2s_2c_1 & c_2s_2s_1 & 2
c^2_2s_1c_1\end{array}\right)
\end{eqnarray}

The neutrino mass matrix ${\cal M}_{\nu}$ has a zero eigenvalue since one
of the three right handed neutrinos was not protected by the $SU(2)_H$
symmetry and had decoupled.  Identifying this with the third neutrino we
have its mass $m_3=0$.  The correspnding third-neutrino eigenvector  is
easy to evaluate exactly and is $(c_2s_1,c_2c_1,-s_2)$.  

The orthogonal matrix that diagonalizes ${\cal M}_{\nu}$
(i.e. $U^{\dagger}_\nu {\cal M}_\nu U_\nu = D_\nu$) is given by
\begin{eqnarray}
U_{\nu}~=~\left(\begin{array}{ccc} c_1c'-s_1s_2s' & c_1s'+s_1s_2c' &
c_2s_1 \\
-s_1c'-c_1s_2s' & -s_1s'+c_1s_2c' & c_2c_1 \\ -c_2s' & c_2c' & -s_2
\end{array}\right)
\end{eqnarray}
where $s'= sin\theta'$ with $\theta'$ given by
$tan2\theta'=\frac{2s_2(c^2_1-s^2_1)}{(1+s^2_2)2s_1c_1}$.
 Note that as $s_1\rightarrow 0$, $\theta'\rightarrow \pi/4$. Note that
the third neutrino eigenvector is the third column of the above matrix.

 The final physical neutrino mixing matrix is then given by ${\bf
\Large U} =U^{(L)}_{\ell}U_{\nu}$, where $U^{(L)}_\ell$ is defined
in the previous section.
Combining this with the neutrino mixing matrix $U_{\nu}$, we get the final
physical neutrino mixing matrix $U$ to be
\begin{eqnarray}
{\bf \Large U}=\left(\begin{array}{ccc} s_2s' & s_2c' & c_2 \\
-(c'c_\beta+c_2s's_\beta) & -(c_\beta s'-s_\beta c_2c') & -s_2s_\beta \\
(c's_\beta - c_2 s' c_\beta) & ( c_2c'c_\beta+s's_\beta) & s_2 c_\beta
 \end{array}\right)
\end{eqnarray}
 To see the consistency of the model, we first note that
$U_{e3}=c_2 \leq 0.16$ from the reactor neutrino data\cite{chooz}. This
implies $s_2\simeq 1$ and for $\kappa_1 \ll \kappa_2 \ll
\kappa_0$, $\frac{h_1\kappa_2}{h_0\kappa_0}\leq 0.16$.
 To fit the atmospheric data, we then require, $s_\beta\simeq
1/\sqrt{2}$. This can be easily satisfied by requiring the Yukawa
couplings to have a hierarchy $h'_4/h'_1 \ll
2\sqrt{\frac{m_{\mu}}{m_{\tau}}}$.

The solar mixing angle $(sin \theta_{\odot}\equiv U_{e2})$, for $s_2\simeq
1$ is given by 
\begin{eqnarray}
tan 2\theta_{\odot}\equiv tan 2\theta' \simeq cot 2\theta_1
\end{eqnarray}
This implies that 
\begin{eqnarray}
sin 2\theta_{\odot} \simeq
(c^2_1-s^2_1) =\frac{\kappa^2_2-\kappa^2_1}{\kappa^2_2 + \kappa^2_1}
\end{eqnarray}


Coming to neutrino masses $m_1, m_2$ and $m_3$, in the absence of
$\kappa_1\kappa_2$ we have $m_1 = -m_2 = h_0\kappa_0$ and $m_3=0$. The
solar mass
squared difference $\Delta m^2_{\odot} = m^2_1 - m^2_2$ is generated when
$\kappa_1$ is turned on while the atmospheric mass squared difference
$\Delta m^2_A \equiv m^2_1 - m^2_2 \simeq h^2_0\kappa^2_0$ gets a small
correction. In the limit of $s_2\simeq 1$, one can write $\Delta
m^2_{\odot}$ in terms
of $\kappa_i$ as follows:
\begin{eqnarray}
\frac{\Delta m^2_{\odot}}{\Delta m^2_A}= \frac{4h^2_1
\kappa_1\kappa_2}{\sqrt{h^2_0\kappa^2_0 + h^2_1(\kappa^2_1 +\kappa^2_2)}}
\end{eqnarray}
It is then clear that if we choose $\kappa_{1,2}\ll \kappa_0$ along with a
mild hierarchy among the Yukawa parameters $h_{1,0}$, we can obtain the
desired solar neutrino mass squared difference. This leaves the relative
valus of $\kappa_1$ with respect to $\kappa_2$ unaffected. Appropriately
choosing their relative values, we can  get the solar mixing angle to be
smaller than maximal as indicated by the central value for it.

Combining the above equations, we get for $U_{e3}\ll 1$,
\begin{eqnarray}
U^2_{e3} cos 2\theta_{\odot} = \frac{\Delta m^2_{\odot}}{ 2 \Delta m^2_A}
+
O(U^4_{e3}, (m_e/m_\mu)^2)
\end{eqnarray}
This equation is the major result of the paper and it is a direct
consequence of the $SU(2)_H$ symmetry. It is interesting to note that the
smallness of $\Delta m^2_{\odot}/\Delta m^2_A$ is related to the smallness
of $U_{e3}$. Furthermore this relation, Eq.(20) provides a test of the
leptonic horizontal symmetry.  In Fig. 1, we show the implications
of this equation for the allowed parameter range in the case of the LMA
solution to the solar neutrino problem. Inside of the quadrilateral is the
allowed region for the parameters for the centarl value of the 
 $\Delta m^2_A = 2.5 \times 10^{-3}$
eV$^2$. Therefore, unless new solar neutrino data changes the current
picture of neutrino mixings, long base line
experiments such as the proposed JHF and NUMI Off-axis\cite{para} as well
as KAMLAND experiments must yield points inside this
allowed region, if this model is to describe nature.
As is clear from the Fig. 1, taking the best fit values for the solar
mixing angle
i.e. $0.22 \leq tan^2\theta_{\odot} \leq 0.59$ and $2.2\times
10^{-5} eV^2 \leq \Delta m^2_{\odot}\leq 2\times 10^{-4} eV^2$ (at 99.7\%
c.l.), we find that, $0.25 \leq cos 2\theta_{\odot} \leq 0.63$, and using
the above equation we get, 
\begin{eqnarray}
 0.4 \geq U_{e3} \geq 0.083
\end{eqnarray}
Thus this model is testable in near future.
\begin{figure}[htb] 
\begin{center}
\epsfxsize=20cm
\epsfysize=20cm
\mbox{\hskip -.50in}\epsfbox{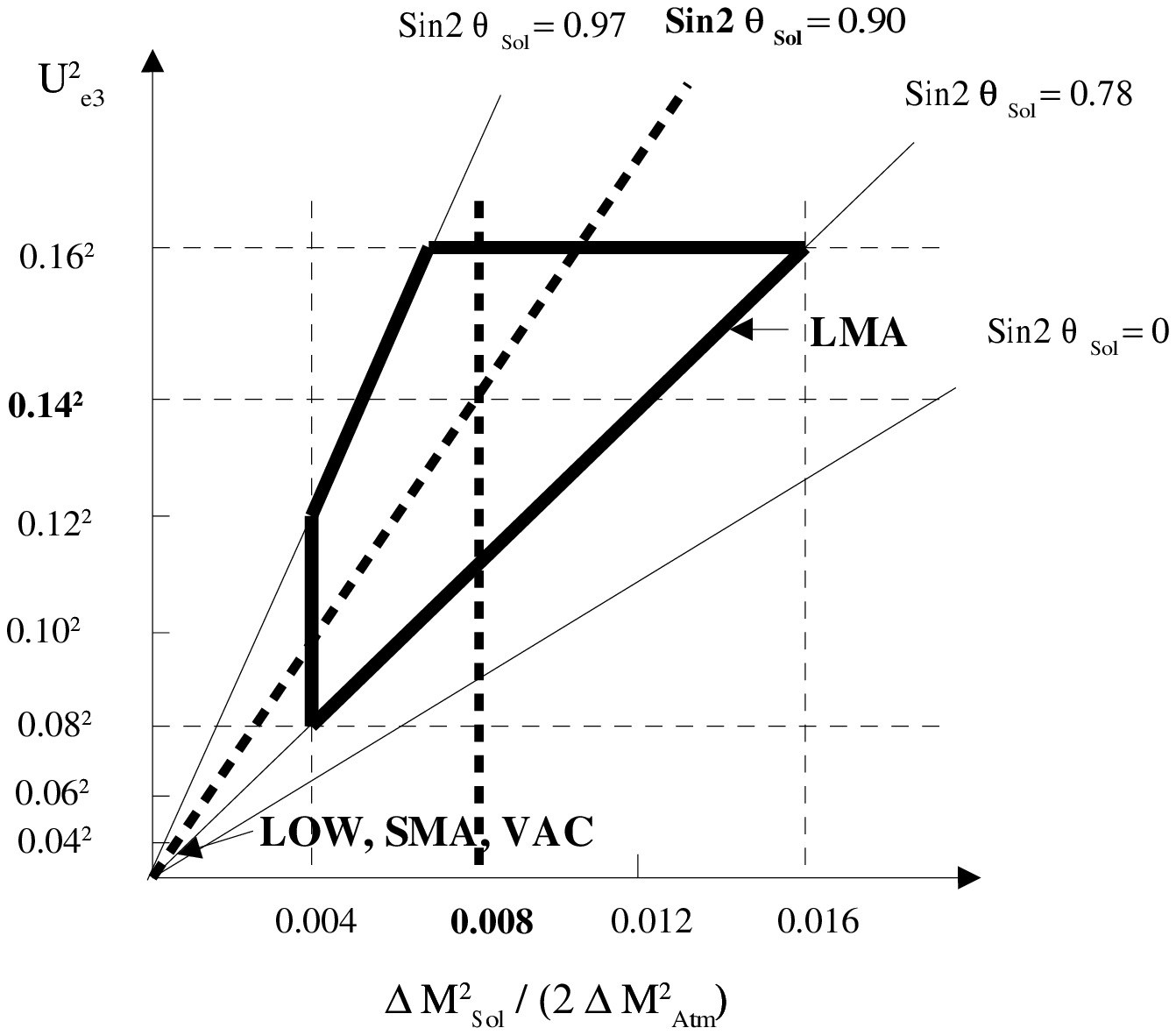}
\caption{
The figure depicts the prediction of our model for $U_{e3}$ for
different values of $\Delta m^2_{\odot}$ and $sin^22\theta_{\odot}$. The
points inside the quadrilateral region are the model predictions. Bold
dotted lines are the current central values for $sin 2\theta_{\odot}$
and $\Delta m^2_{\odot}$. The labels SMA, LOW and VAC mean the location of
the relevant solutions in the plot.
\label{Fig.1}}
\end{center}
\end{figure}

\section{Comments and other tests of the model}

(i)  A characteristic test of the inverted hierarchy models is in its
prediction for neutrinoless double beta decay\cite{smirnov}. Generically
in the exact bimaximal limit, the effective mass measured in neutrinoless
double beta decay i.e. $< m>_{\beta\beta}\equiv \sum U^2_{ei} m_i = 0$,
since we have $< m>_{\beta\beta}= m_1 U^2_{e1}+m_2U^2_{e2}$ and
$m_1=-m_2$ and $U_{e1}=U_{e2} = 1/\sqrt{2}$.
The mass matrix in Eq. (8) however differs from this limit; nonetheless,
the change in mass differences and the change in the mixing angles
compensate each other to give zero. Therefore in the class of models we
are discussing, the neutrinoless double beta decay is a probe of the
structure of the leptonic mass matrix.
In our case we predict $< m>_{\beta\beta}\simeq (cos
2\theta_{\odot}\sqrt{\Delta m^2_A})$, which at 99.7\% confidence level be
as large as $0.03$ eV.

(ii) The model leads to the standard MSSM below the horizontal symmetry
breaking scale.

(iii) In our model we imposed  $\bar{\chi} \rightarrow -\bar{\chi}$
discrete symmetry.  This will leave the charge neutral Higgisnos
corresponding to $\chi$ and $\bar{\chi}$ massless.  However one may add a
bilinear term of the form $\bar{\chi}\chi$ to the superpotential thereby
breaking the discrete symmetry softly. This will then generate a mass for
the corresponding horizontal Higgsino.

\section{Summary and conclusions}
In conclusion, we have presented an extension of the minimal
supersymmetric standard model where
a local $SU(2)_H$ symmetry acts on both on the left and right handed 
charged leptons. Freedom from
global anomalies then requires that a doublet of right handed neutrinos
be included in the theory. This model provides a very natural way to
understand two crucial features of the current neutrino oscillation data
i.e. near bimaximal mixing pattern and a small $\Delta m^2_{\odot}/\Delta
m^2_A$. It also gives a relation between the neutrino observables
$U_{e3}$, $\Delta m^2_{\odot}/\Delta m^2_A$ and solar mixing angle
$sin^22\theta_{\odot}$. For the current fits to the latter two parameters,
it predicts a lower bound on $U_{e3}$ which is quite accessible to long
baseline experiments currently planned. We also present the complete range
of allowed values for $U_{e3}$ and $\Delta m^2_{\odot}/\Delta m^2_A$
predicted by our model.

\bigskip

The work of R. N. M. is supported by the National Science Foundation Grant
No. PHY-0099544. We thank Luis Lavoura for a useful comment and Dragos
Constantin for help with the figure.

\end{document}